# Basic Physical Properties of Cubic Boron Arsenide

**Joon Sang Kang, Man Li, Huan Wu, Huuduy Nguyen, Yongjie Hu\***

School of Engineering and Applied Science, University of California, Los Angeles (UCLA),

Los Angeles, California 90095, United States

\*Corresponding author. Email: yhu@seas.ucla.edu

**Abstract**

Cubic boron arsenide (BAs) is an emerging semiconductor material with a record-high thermal conductivity subject to intensive research interest for its applications in electronics thermal management. However, many fundamental properties of BAs remain unexplored experimentally since high-quality BAs single crystals have only been obtained very recently. Here, we report the systematic experimental measurements of important physical properties of BAs, including the band gap, optical refractive index, elastic modulus, shear modulus, Poisson's ratio, thermal expansion coefficient, and heat capacity. In particular, light absorption and Fabry-Perot interference were used to measure an optical band gap of 1.82 eV and a refractive index of 3.29 (657 nm) at room temperature. A picoultrasonics method, based on ultrafast optical pump probe spectroscopy, was used to measure a high elastic modulus of 326 GPa, which is twice that of silicon. Furthermore, temperature-dependent X-ray diffraction was used to measure a linear thermal expansion coefficient of $3.85 \times 10^{-6}$ K$^{-1}$; this value is very close to prototyped semiconductors such as GaN, which underscores the promise of BAs for cooling high power and high frequency electronics. We also performed *ab initio* theory calculations and observed good agreement between the experimental and theoretical results. Importantly, this work aims to build a database (Table 1) for the basic physical properties of BAs with the expectation that this semiconductor will inspire broad research and applications in electronics, photonics, and mechanics.

Heat dissipation is a critical technology issue for modern electronics and photonics[1-5]. A key challenge and urgent need for effective thermal management is to discover new materials with ultrahigh thermal conductivity for dissipating heat from hot spots efficiently, and thereby improving device performance and reliability[6-10]. Options to address this challenge with conventional high thermal conductivity materials, such as diamond and cubic boron nitride, are very limited[11-13]; their high cost, slow growth rate, degraded crystal quality, combined with the integration challenge with semiconductors, given chemical inertness and large mismatch in lattice and thermal expansion, make them far from optimal candidates. Carbon-based nanomaterials such as graphene and carbon nanotube, are limited by their intrinsic anisotropic thermal conductivity and performance degradation due to ambient scattering despite that individual materials can have high thermal conductivity[14]. Recent *ab initio* theory calculation has predicted new boron compounds with high thermal conductivity stemming from the fundamental vibrational spectra[15-17]. Although the early growth was reported in 1950s[18-23], obtaining high-quality crystals of these materials with minimal defects has proven challenging and was only achieved very recently[6-9]. Our recent study[7] reports the successful synthesis of single-crystal BAs with non-detectable defects, along with the measurement of a record-high thermal conductivity of 1300 W/mK, a value which is beyond that of most common semiconductors and metals. Thus, BAs is a promising new semiconductor that could potentially revolutionize the current technology paradigms of thermal management and possibly extend the semiconductor roadmap. However, many of the basic physical properties of BAs have yet to be explored experimentally. Here, we report the systematic experimental study of our chemically synthesized high-quality BAs crystals, including the measurement of basic BAs characteristics such as, the bandgap, optical refractive index, sound velocity, elastic modulus, shear modulus, Poisson's ratio, thermal expansion coefficient, and heat capacity.

High-quality single-crystals BAs (Figure 1) were synthesized by chemical vapor transport reaction as described in our previous work[7]. BAs has a zinc-blended face-centered crystal structure with a F43m space group, where boron and arsenic atoms are interpenetrating each other and covalently bonded with a lattice constant of 4.78 Å. The synthetic BAs crystals, characterized with a high-resolution scanning transmission electron microscopy, show an atomically resolved lattice.

We report the experimental measurement of the optical bandgap ($E_g$) of BAs. $E_g$ is a major factor that determines the optical absorption and electrical transport in a solid. However, owing to the synthesis challenge to obtain high-quality crystals until recently[7], the $E_g$ of bulk BAs remained inconclusive with theoretical calculated values varying from 0.67 to 5.5 eV[24–30]. Here we measured the light absorption of BAs single crystals to determine $E_g$ using ultraviolet–visible spectroscopy. The transmission spectrum of BAs was collected by scanning light wavelengths with photon energy from 1.5 to 2.2 eV, following the Beer–Lambert law:

$$I/I_0 = e^{-\alpha d} \qquad (1)$$

where *I* and $I_0$ are the transmitted light and the intensity of incident light, respectively, and α and *d* are the absorption coefficient and the thickness of sample crystal, respectively. Figure 2a shows the absorption measurement data. As an indirect band gap semiconductor, the band edge absorption of BAs was determined by[31]:

$$\alpha \propto (h\nu - E_g \pm \hbar\Omega)^2 \qquad (2)$$

where *h, and v* are respectively the Planck constant and light frequency. Equation (2) indicates a threshold in the absorption close to $E_g$, but with a small $\hbar\Omega$ difference due to the contribution of phonons during the light absorption process, as required by momentum conservation. The absorption data plot of $\alpha^{\frac{1}{2}}$ versus $h\nu$ fits well onto a straight line, which confirms Equation (2) and enables the extrapolation of $E_g$ ~ 1.82 eV. Note that $E_g$ could be slightly shifted due to the absorption or emission of phonons, whose energies at the X point are 0.02 ~ 0.04 eV for transverse acoustic (TA) and longitudinal acoustic (LA) branches. In addition, the experimental curve shows a tail extending down to about 1.5 eV, which is caused by the absorption of high-energy phonons or multi-phonons.

We also measured the refractive index of BAs. The refractive index (n) is a key dimensionless parameter for photonic applications that describes how fast light propagates through the material, but was unknown for BAs so far. Theoretical refractive indexes of BAs appearing in recent literature have different values, possibly due to variations in pseudo-potentials[28,30,32] or empirical relationships[33]. Here, we directly measured n by Fabry-Perot interference. When light passes through a BAs crystal (inset of Figure 2b), the top and bottom surfaces can serve as two parallel reflecting mirrors. The transmission spectrum measured as a function of wavelength from 657 to 908 nm exhibits large resonance oscillations (Figure 2b). *n* can be extracted from following equation under normal light incidence[34,35]:

$$n = \frac{M\lambda_1\lambda_2}{2(\lambda_1-\lambda_2)d} \quad (3)$$

where M is the number of fringes between the two extrema (M=1 for consecutive maxima), and $\lambda_1$ and $\lambda_2$ are the corresponding light wavelength between fringes. The refractive index of BAs was measured as n= 3.29 to 3.04 with increased optical wavelengths from 657 to 908 nm and plotted in Figure 2c, which is close to that of prototyped optoelectronic materials, such as GaP (3.2 at 900 nm)[36] and GaAs (3.7 at 876 nm)[37].

Next we measured the basic mechanical properties of BAs, including elastic modulus, shear modulus, sound velocity, and Poisson's ratio. The stresses and strains of a material are connected by Hooke's spring law:

$$\sigma_{ij} = C_{ijkl}\varepsilon_{kl} \qquad (4)$$

where σ, C, and ε are the stress, stiffness, and strain, respectively. In general, for anisotropic materials, $C_{ijkl}$ is a fourth-order tensor with 81 components[38,39]. Under the symmetry and Voigt notation[40], the stiffness tensor can be simplified to 36 components. For cubic symmetry materials, such as BAs, only 3 constitutive components— $C_{11}$, $C_{12}$, $C_{44}$— are independent. They are closely related with the sound velocities as discussed below.

To measure the stiffness of BAs, we first performed picosecond ultrasonic measurements[41] to determine sound velocity using a pump-probe ultrafast optical spectroscopy[42–45] (Figure 3a). In this setup, a thin metal layer (aluminum) is deposited on BAs surface to serve as the optical transducer and a picosecond laser pump generates a longitudinal acoustic wave that penetrates into the BAs sample, travels across it, reflects back, and creates sound echoes. The sound echoes are detected by a probe pulse laser with a controlled sub-picosecond stage delay (Figure 3b). The time delay ($\Delta t_{echo}$) between echoes is measured and contributed to by the round trip of the acoustic wave in the BAs film with a given thickness (d), and thus determining the sound velocity,

$$v_{LA} = \frac{2 \cdot d}{\Delta t_{echo}} \qquad (5)$$

For example, we prepared a BAs thin film slab along the (111) orientation using focused ion beam (FIB) (inset of Figure 3b) and the room temperature $v_{LA,111}$ is 8150 ± 450 m/s. We also measured the temperature dependence from 80 to 500 K (Figure 3c), indicating a slightly decrease in sound velocity with increased temperature. Note that the stiffness $C_{44}$ can be linked with $v_{LA,111}$, $C_{11}$ and $C_{12}$ with the following formula[46],

$$C_{44} = \frac{(3\rho v_{LA,111}^2 - C_{11} - 2C_{12})}{4} \quad (6)$$

where the mass density ρ is 5.22 g/cm³ for BAs[7]. Since $C_{11}$ and $C_{12}$ can be linked by[38] $C_{12} = \frac{3B-C_{11}}{2}$ where B is the bulk modulus reported as 148 GPa[47,48], $C_{11}$ cancels out in Equation (6) leaving a relationship between $C_{44}$ and $v_{LA,111}$. Therefore, from these results, the stiffness $C_{44}$ is determined to be 149 GPa. The transverse acoustic velocity along <100> is $v_{TA} = \sqrt{\frac{C_{44}}{\rho}} = 5340 \pm 510$ m/s.

To obtain the stiffness $C_{11}$, we measured surface acoustic waves (SAW)[49,50] (Figure 3d). Periodic metal lines (with a period of L) were fabricated on the surface of BAs samples in alignment with the desired crystal directions using electron beam lithography[51], and heated up by a pulsed laser to generate periodic thermal strains that partially pass through the sample to form SAWs (Figure 3d)[44,52,53]. In fact, due to the cubic structure of BAs, the in-plane acoustic velocities for the (111) crystal surface can be considered as isotropic. The SAW propagating along the BAs surface and detected by a laser probe shows strong periodic oscillations with a time periodicity of $\Delta t_{SAW}$ in the ultrafast time domain (Figure 3e), which measures the SAW velocity. For example, for the main top crystal surface in the (111) orientation, the SAW velocity was measured as: $v_{SAW} = L/\Delta t_{SAW} = 4320 \pm 120$ m/s, which is used to extract the $C_{11}$ of BAs as discussed below.

In general, the bulk wave equation for the displacement in a perfectly elastic and anisotropic medium can be written

$$C_{ijkl} \frac{\partial^2 u_k}{\partial x_l \partial x_j} = \rho \frac{\partial^2 u_i}{\partial t^2} \quad (7)$$

where the $u_i$ are the displacement components along the Cartesian axes $x_i$ to which the stiffness tensor $C_{ijkl}$ is referred. $\rho$ is the mass density and the Einstein summation is implied. For bulk waves, the solutions of Equation (7) are plane waves. However, for SAWs, these solutions represent the progagation modes of elastic energy along the free surface of a half-space, and their displacement amplitudes decay exponentially with depth, so that most energy is concentrated within a wavelength distance below the surface[50,54]. When a combination coefficient (*A*) for each wave component is used to consider the direction-dependent free propagation, space confinement, and amplitude decays respectively, the solution for SAWs can be written as linear combinations[50]

$$u_i = \sum_m A^{(m)} U_j^{(m)} \exp(ik(l_i x_i - v_{SAW} \cdot t)) \quad (8)$$

where *k*, $U_j$, $\delta_{ij}$, *l* are the wave vector, amplitude, Kronecker delta, and directional vector respectively. The solution of equation (8) can be determined by substituting the wave forms of equation (8) into the equation (7) that leads to the Christoffel equation

$$[k^2 l_i l_l c_{ijkl} - \rho \omega^2 \delta_{ij}][U_i] = 0 \quad (9)$$

where ω is the SAW frequency, ω = $k \cdot v_{SAW}$. Thus, the only unknown parameter $A$ can be determined by using the periodic mass boundary condition[55] of the experimental system and the eigenvector $l_i$ and $U_i$ can be obtained by solving Equations (7-9)[50]. The stiffness $C_{11}$ is thereby determined to be 285 GPa and the longitudinal velocity in the [100] direction, calculated by $v_{LA} = \sqrt{\frac{C_{11}}{\rho}}$, is 7390 m/s.

Therefore, all the stiffnesses of BAs were measured as $C_{11}$ = 285 GPa; $C_{12}$ = 79.5 GPa; and $C_{44}$ = 149 GPa. The elastic modulus of an arbitrary direction is determined from the elastic compliances[56]

$$E_{hkl} = \frac{1}{S_{11} - 2(S_{11} - S_{12} - 0.5 S_{11})(a_1^2 a_2^2 + a_2^2 a_3^2 + a_3^2 a_1^2)} \quad (10)$$

where <hkl> is the Miller index and $a_1$, $a_2$, $a_3$ are directional cosine. Hence, $E_{100}$, $E_{110}$, and $E_{111}$ are 250, 308, and 335 GPa, respectively. In addition, the Poisson's ratio of BAs, determined similarly to the equation (10), is[56]

$$v_{ab} = - \frac{S_{12} + (S_{11} - S_{12} - 0.5 S_{44})(e_1^2 b_1^2 + e_2^2 b_2^2 + e_3^2 b_3^2)}{S_{11} - 2(S_{11} - S_{12} - 0.5 S_{44})(e_1^2 e_2^2 + e_2^2 e_3^2 + e_3^2 e_1^2)} \quad (11)$$

where $e_i$ and $b_i$ are directional cosine of the transversal and axial strain. The Poisson's ratio in the <100> on the (100) plane was measured as 0.22. Additionally, the averaged elastic ($E_{VRH}$) and shear moduli ($G_{VRH}$) of BAs can be determined by Voigt-Reuss-Hill average method[57]

$$E_{VRH} = \frac{1}{2}\left(\frac{(C_{11} - C_{12} + 3C_{44})(C_{11} + 3C_{12})}{2C_{11} + 3C_{12} + C_{44}} + \frac{5}{3S_{11} + 2S_{12} + S_{44}}\right) \quad (12)$$

$$G_{VRH} = \frac{1}{2}\left(\left(\frac{C_{11} - C_{12} + 3C_{44}}{5}\right) + \left(\frac{5}{4(S_{11} - S_{12}) + 3S_{44}}\right)\right) \quad (13)$$

The averaged elastic and shear moduli were measured as 326 and 128 GPa, respectively. Despite a lower sound velocity, it should be noted that the $E_{VRH}$ is about twice as high as that of silicon (~160 GPa) and comparable to that of GaN (~300 GPa), due to the large mass density. This result suggests that BAs is a good mechanical material for micromachining and high pressure applications.

To analyze the experimental results, we applied *ab initio* theory to calculate the mechanical properties of BAs including the bulk modulus, elastic modulus, and the Poisson's ratio. We applied the density functional theory (DFT) to calculate the force acting on each atom using Quantum ESPRESSO package[58,59]. We used projection-augmented wave pseudopotentials under local density approximation for both boron and arsenic[60]. An 8 atoms unit cell with 6 × 6 × 6 Monkhorst-Pack k-points meshes is used for relaxation, and the relaxed lattice constant for BAs is 4.7434 Angstrom. We expand the lattice constants by 0.001 Angstrom, and calculate the pressure applied on the unit cell by DFT. The calculated bulk modulus, $C_{11}$, $C_{12}$, $C_{44}$, and Poison's ratio (<100>) are respectively 150 GPa, 294 GPa, 80.6 GPa, 177 GPa, and 0.22, in good agreement with experiments.

We also measured the thermal expansion coefficients of BAs by temperature dependent X-ray diffraction (XRD) using a Cu Kα radiation source. Figure 4a shows the XRD data from 293 to 773 K. As the temperature increases, the (311) peak clearly shifts to a smaller angle (Figure 4a), indicating a decrease in the d-space and an increase of the lattice constant. The measured lattice constant of BAs at 298K is 4.78Å and agrees with our previous report[7]. The temperature-dependent lattice constants (*l*) of BAs (Figure 4b) were used to obtain the linear thermal expansion coefficient ($\alpha_L$),

$$\alpha_L \equiv \frac{1}{l}\frac{dl}{dT} \qquad (14)$$

The measured $\alpha_L$(T) is fit using the least square method with a second-order polynomial curve (Figure 4b) to obtain the thermal expansion coefficient in Figure 4c: $\alpha_L$ is 3.85×10$^{-6}$/K at 298 K and increases to 6.06×10$^{-6}$/K at 723 K. Due to its cubic structure, BAs has a volumetric thermal expansion $\alpha_v = 3 \cdot \alpha_L$ = 11.55×10$^{-6}$/K at room temperature. Here, we also performed *ab initio* calculations of the thermal expansion[63–65]. The thermal expansion coefficient can be calculated from phonon dispersion and mode Grüneisen parameters. The phonon dispersion is calculated from the 2$^{nd}$ order interatomic force constants (IFCs), and Grüneisen parameters are calculated from the 2$^{nd}$ and 3$^{rd}$ order IFCs using ALAMODE package[66] based on 216 atoms supercell using finite displacement plus least square fitting method. Our measurement results shows good agreement with the calculation (Figure 4c) and recent study[61,62].

Knowledge of the thermal expansion property of BAs is critical for its integration into high-power electronics materials to take advantage of the ultrahigh thermal conductivity. A thermal expansion match between device layers is desired to minimize the residual stress and prevent the delamination of materials with their adjacent layers, as well as to reduce the thermal boundary resistance[67]. Therefore it is worthy to compare $\alpha_L$ of BAs with those of GaN and GaAs, as they are potential target materials for passive cooling application of BAs. $\alpha_L$ of GaN is 3.94×10$^{-6}$/K at 300 K and 5.33×10$^{-6}$/K at 800K (a-axis)[68], and that of GaAs is 5.9×10$^{-6}$/K at 300K[69]. These values are very close to our measured thermal expansion coefficients of BAs, thereby underscoring the high promise of BAs as a superior cooling material for high power and high frequency GaN or GaAs electronic devices. In comparison, those prototype high thermal conductivity materials diamond (1.6×10$^{-6}$/K)[70] and SiC (3.2×10$^{-6}$/K)[71] at room temperature, have a much larger thermal expansion mismatch that can potentially lead to thermal mechanical instability and large thermal boundary resistances.

We also measured the temperature dependent specific heat ($C_v$) of BAs using differential scanning calorimetry in Figure 4d. In addition, we performed *ab initio* calculations[72] using the same DFT settings. When plotted together, our experimental data from this work (solid squares) and prior work[7,73](open circles) show good agreement with the theory (Figure 4d). Moreover, the Grüneisen parameter ($\gamma$ )[74] is generally used to describes the effect of volumetric change of a crystal on the vibrational properties , and is a weighted average overs all phonon modes[45] and can be calculated by

$$\gamma = \frac{\alpha_v B}{C_v} \qquad (15)$$

where $\alpha_v$ and $C_v$ are the volume expansion coefficient and specific heat capacity, respectively. In our study, the room temperature Grüneisen parameter of BAs was determined as $\gamma = 0.82$.

In summary, we report systematic experimental measurements of important physical properties of cubic BAs, including the band gap, refractive index, elastic modulus, shear modulus, Poisson's ratio, and thermal expansion coefficient, and specific heat capacity. The database of the physical properties of BAs, summarized in Table 1, is now complete after addition of the experimental values from this work to the literature. With an ultrahigh thermal conductivity of 1300 W/mK[7], we believe that the systematically reported physical properties of BAs will further facilitate the development of this new semiconductor for thermal, electrical, optical, and mechanical applications.

## Acknowledgements

Y. H. acknowledges support from a CAREER award from the National Science Foundation under grant DMR-1753393, an Alfred P. Sloan Research Fellowship under grant FG-2019-11788, a Young Investigator Award from the US Air Force Office of Scientific Research under grant FA9550-17-1-0149, a Doctoral New Investigator Award from the American Chemical Society Petroleum Research Fund under grant 58206-DNI5, as well as from the UCLA Sustainable LA Grand Challenge and the Anthony and Jeanne Pritzker Family Foundation.

**Table 1. Summary of experimentally measured physical properties of BAs.**

| Physical property | Experimental value | References |
|---|---|---|
| Crystal structure | Zinc-blende cubic ($F\bar{4}3m$) | Ref. 7-9, 18 |
| Lattice constant (Å) | 4.78 | Ref. 7, 18 |
| Band gap (eV) | 1.82 | This work |
| Refractive index | 3.29 (657nm) <br> 3.04 (908nm) | This work |
| Mass density (g/cm$^3$) | 5.22 | Ref. 7, 18 |
| Stiffness $C_{11}$, $C_{12}$, $C_{44}$ (GPa) | 285, 79.5, 149 | This work |
| Compliance $S_{11}$, $S_{12}$, $S_{44}$ ($\times 10^{-12}$ Pa$^{-1}$) | 3.99, -0.87, 6.71 | This work |
| Averaged elastic modulus (GPa) | 326 | This work |
| Averaged shear modulus (GPa) | 128 | This work |
| Bulk modulus (GPa) | 148 | Ref. 47 |
| Poisson's ratio | 0.22 (<100> on (100)) | This work |
| Longitudinal sound velocity (m/s) | 7390 (<100>) <br> 8150 (<111>) | This work |
| Transverse sound velocity (m/s) | 5340 (<100>) | This work |
| Thermal conductivity (W/m·K) | 1300 | Ref. 7 |
| Volumetric heat capacity (J/cm$^3$ K) | 2.09 | This work, Ref. 7, 73 |
| Thermal expansion coefficient (10$^{-6}$ K$^{-1}$) | 3.85 (linear) <br> 11.55 (volume) | This work |
| Grüneisen parameter | 0.82 | This work |

# Figure captions

**Figure 1**. **Cubic BAs crystals.** a) Unit cell structure of BAs; b) High-resolution scanning transmission electron microscopy image of BAs with <110> zone axis. Inset, optical image of a BAs thin film sample.

**Figure 2. Optical properties of BAs.** a) Band gap measurement from light absorption experiments. The absorptivity is plotted versus the photon energy of a 55 µm thick BAs sample; (b) Refractive index measurement from a Fabry-Perot interference experiment: the optical transmission measured on a 8.79 µm thick BAs sample shows strong oscillations due to resonances from the light reflected by the top and bottom sample surfaces (inset); and c) Wavelength dependent refractive index of BAs.

**Figure 3**. **Mechanical properties of BAs.** a) Schematic of the picosecond ultrasonic measurement process. b) Picosecond acoustic data: the sound echoes indicate a round trip of the acoustic wave generated inside the BAs sample. The inset shows a thin film (4.37 µm) of BAs prepared with a focused ion beam in the (111) plane; c) Temperature dependent sound velocity of BAs; d) Schematic of surface acoustic wave (SAW) measurement using ultrafast pump-probe spectroscopy; e) The detection signal as a function of delay time shows periodic oscillations corresponding to the propagation of a SAW. The inset shows the fabricated metal patterns for SAW measurement using e-beam lithography. The scale bar is 1µm; and f) The measured Poisson's ratio of BAs on the (100) plane (black line) and (111) plane (red line).

**Figure 4. Thermal properties of BAs.** a) The temperature dependent X-ray diffraction (XRD) measurement for (311) peaks with temperatures from 298K to 773 K; b) The measured lattice constant of BAs as a function of temperature. The red line represents the second order polynomial fitting; c) The linear thermal expansion coefficient ($\alpha_L$) of BAs with temperature dependence; and d) The specific heat ($C_v$) of BAs from 5 to 600 K. Symbols indicate experimental data and the red line represents the DFT calculations.

**Table 1. Summary of experimentally measured physical properties of BAs.**

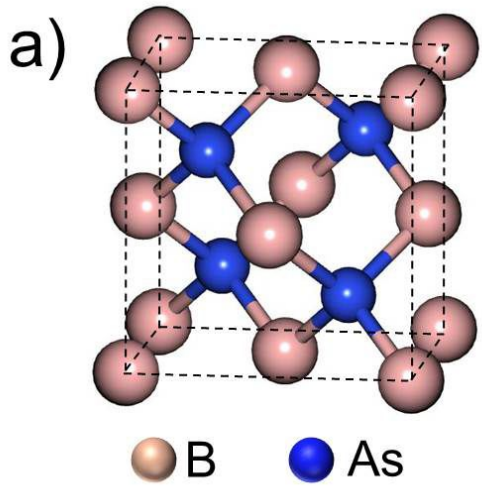 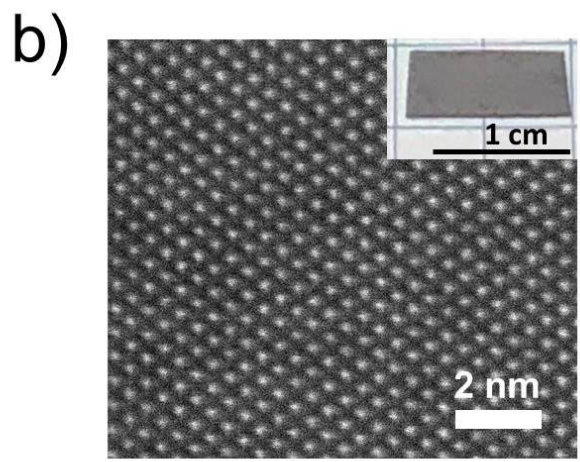

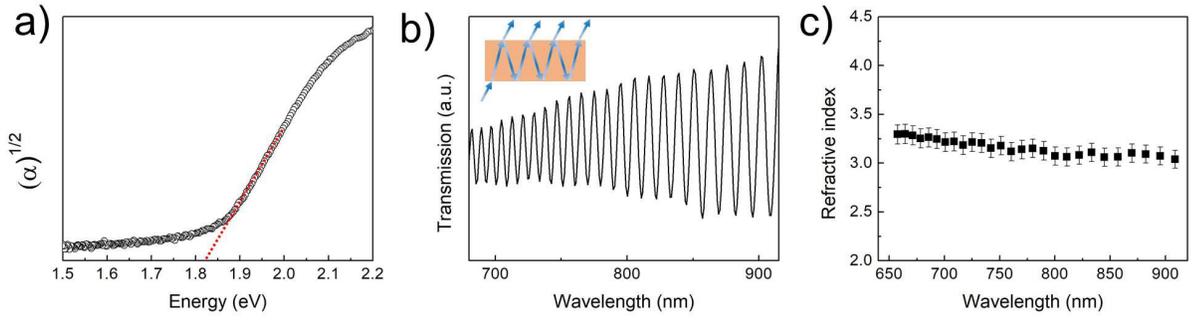

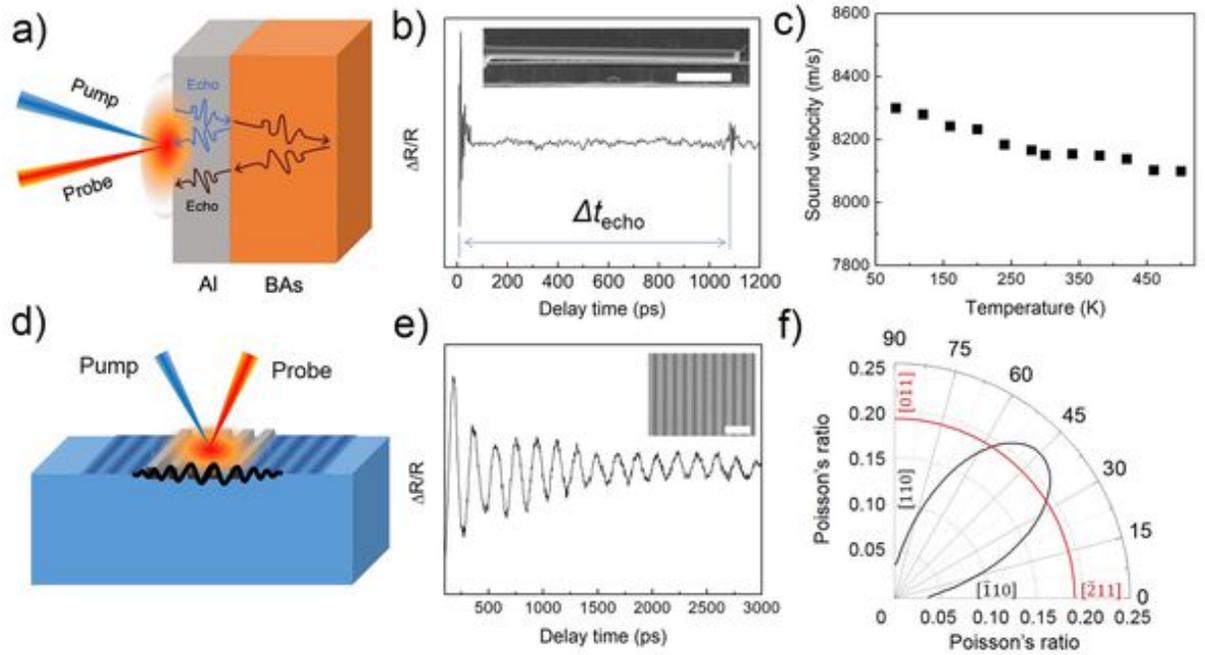

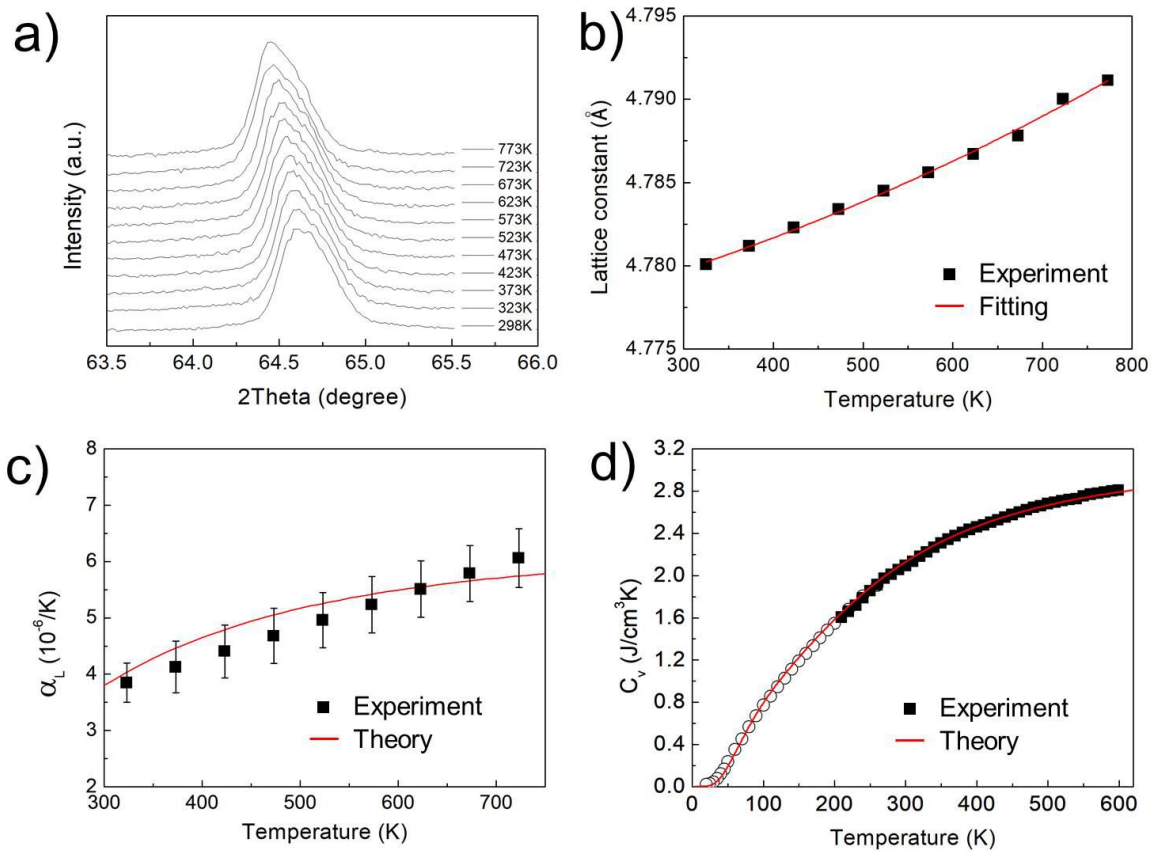

TABLE I. Summary of experimentally measured physical properties of BAs.

| Physical property | Experimental value | Reference |
|---|---|---|
| Crystal structure | Zinc-blende cubic ($F\bar{4}3m$) | Ref. 7-9, 18 |
| Lattice constant (Å) | 4.78 | Ref. 7, 18 |
| Band gap (eV) | 1.82 | This work |
| Refractive index | 3.29 (657nm) 3.04 (908nm) | This work |
| Mass density (g/cm$^3$) | 5.22 | Ref. 7, 18 |
| Stiffness $C_{11}$, $C_{12}$, $C_{44}$ (GPa) | 285, 79.5, 149 | This work |
| Compliance $S_{11}$, $S_{12}$, $S_{44}$ (×10$^{-12}$ Pa$^{-1}$) | 3.99, -0.87, 6.71 | This work |
| Averaged elastic modulus (GPa) | 326 | This work |
| Averaged shear modulus (GPa) | 128 | This work |
| Bulk modulus (GPa) | 148 | Ref. 47 |
| Poisson's ratio | 0.22 (<100>) | This work |
| Longitudinal sound velocity (m/s) | 7390 (<100>) 8150 (<111>) | This work |
| Transverse sound velocity (m/s) | 5340 (<100>) | This work |
| Thermal conductivity (W/m·K) | 1300 | Ref. 7 |
| Volumetric heat capacity (J/cm$^3$ K) | 2.09 | This work, Ref. 7, 73 |
| Thermal expansion coefficient (10$^{-6}$ K$^{-1}$) | 3.85 (linear) 11.55 (volume) | This work |
| Grüneisen parameter | 0.82 | This work |

14